\documentclass[
superscriptaddress,
 amsmath,amssymb,
 aps,
pra,
]{revtex4-2}

\usepackage{graphicx}
\usepackage{dcolumn}
\usepackage{bm}

\usepackage{hyperref}
\hypersetup{
    colorlinks=true,
    citecolor=blue,
    linkcolor=blue,
    filecolor=magenta,      
    urlcolor=magenta,
    pdftitle={Thin-film-flow-over-a-spinning-disc},
    pdfpagemode=FullScreen,
    }

\usepackage{xcolor}
\makeatletter
\providecommand*{\diff}%
{\@ifnextchar^{\DIfF}{\DIfF^{}}}
\def\DIfF^#1{%
	\mathop{\mathrm{\mathstrut d}}%
	\nolimits^{#1}\gobblespace}
\def\gobblespace{%
	\futurelet\diffarg\opspace}
\def\opspace{%
	\let\DiffSpace\!%
	\ifx\diffarg(%
	\let\DiffSpace\relax
	\else
	\ifx\diffarg[%
	\let\DiffSpace\relax
	\else
	\ifx\diffarg\{%
	\let\DiffSpace\relax
	\fi\fi\fi\DiffSpace}

\providecommand*{\pderiv}[3][]{%
	\frac{\partial^{#1}#2}%
	{\partial #3^{#1}}}

\begin{document}

\preprint{APS/123-QED}

\title{Thin film flow over a spinning disc:\\Experiments and direct numerical simulations}

\author{Jason Stafford}
\email{j.stafford@bham.ac.uk}
\affiliation{
 School of Engineering, University of Birmingham, B15 2TT, United Kingdom\\
}%
\affiliation{%
 Department of Chemical Engineering, Imperial College London, SW7 2AZ, United Kingdom\\
}%

\author{Nwachukwu Uzo}
\affiliation{%
 Department of Chemical Engineering, Imperial College London, SW7 2AZ, United Kingdom\\
}%

\author{Enrico Piccoli}
\affiliation{Department of Chemical Engineering, Delft University of Technology, 2629 HZ Delft, The Netherlands\\
}%

\author{Camille Petit}
\affiliation{%
 Department of Chemical Engineering, Imperial College London, SW7 2AZ, United Kingdom\\
}%

\author{Omar K. Matar}
\email{o.matar@imperial.ac.uk}
\affiliation{%
 Department of Chemical Engineering, Imperial College London, SW7 2AZ, United Kingdom\\
}%

\date{\today}

\begin{abstract}
\noindent The dynamics of thin liquid films flowing over a spinning disc is studied through a combination of experiments and direct numerical simulations. We consider a comprehensive range of interfacial flow regimes from waveless through to three-dimensional (3D) waves, and for previously unexplored inertia-dominated conditions that have practical relevance. The transition between these regimes is categorised within a phase map based on two governing parameters that correspond to modified inverse Weber ($\lambda$) and Ekman numbers ($r_{disc}$). Our findings show that stationary two-dimensional (2D) spiral waves, which unfold in the direction of rotation from the Coriolis effect, transition to 3D waves with the emergence of small perturbations on the wavefronts. These non-stationary structures grow asymmetrically in the 2D-3D transitional region, and detach from the parent spiral wave to form wavelets or so-called $\Lambda$ solitons. We show that during and after this wave formation process, flow circulations unique to the spinning disc arrangement are present within the main wave hump. Furthermore, when combined with observations of wall strain rates and topology within the film, these findings elucidate the mechanisms that underpin the apparent wave-induced interfacial turbulence effects observed for spinning disc flows.  
\end{abstract}

\maketitle


\section{\label{sec:Introduction} Introduction}

\noindent The transport of thin liquid films over surfaces is ubiquitous in numerous applications including chemical reactors \cite{Boodhoo2000a,Boodhoo2000b}, spin coating processes \cite{Bai2024}, heat exchangers and evaporators \cite{Alhusseini1998}, and shear-driven synthesis of advanced materials \cite{Stafford2021}. A common feature of viscous liquid films in many of these applications is the formation of surface waves which provide benefits by promoting heat and mass transport over waveless flows, described by the Nusselt solution \cite{Craster2009,Albert2014,Alhusseini1998}. This has generated a wide interest into the dynamics of thin films and surface waves, with some of the earliest experimental and theoretical works by Kapitza on gravity-driven falling films \cite{Kapitza1948,Kapitza1949}. Here, falling films were shown to exhibit a wide variety of nonlinear phenomena, from solitary waves, which evolve downstream to contain subsidiary wavefronts, to 
three-dimensional (3D) instabilities, and finally irregular fully-developed waves \cite{Kapitza1949,Liu_Paul_Gollub_1993}.  

Since then, significant progress has been made on the development of low-dimensional models that predict surface wave instabilities for falling films in two and three dimensions, and including non-isothermal, Marangoni, topographical, and flexible substrate effects \cite{Kalliadasis2013,Bielarz2003,Matar2007}. Several of these methods have been used to investigate solitary waves for varied Reynolds number, Weber number, inclination angle, and periodic flow forcing. This has permitted comparisons with experiments that control the frequency and amplitude of inlet perturbations, parameters which have been found to strongly influence the nonlinear evolution of waves \cite{Liu_Paul_Gollub_1993}: from sinusoidal to solitary, so called $\gamma_1$ and $\gamma_2$ wave families, respectively. Investigations have covered different regimes of falling films, including drag-gravity regimes -- where viscous and gravitational effects are important -- to drag-inertia regimes, where a steep wavefront and long tail emerges from the balance of viscous, gravitational, surface tension, and inertial effects \cite{Rohlfs2015}. Based on the Shkadov scaling for reduced Reynolds number, $\delta = (3Re)^{11/9}\mathit{\Gamma}^{-1/3}$ \cite{Shkadov1977,Kalliadasis2013}, in which $\Gamma$ is the Kapitza number and $Re$ is the film Reynolds number, the transition between these regimes was shown to occur when $\delta \sim 1$ and typically $\delta < 6$ \cite{Rohlfs2015}.    

Over the past three decades, research into thin film flows, particularly falling films, has benefited greatly from advancements in computational resources and experimental techniques. Chakraborty et al. \cite{Chakraborty2014} showed that limitations exist in the use of low-dimensional models when inertia becomes important ($\delta \approx 1$). Results from two-dimensional (2D) direct numerical simulations (DNS) of a solitary, $\gamma_2$ wave were performed across part of the drag-inertia regime ($1<\delta<10$), demonstrating quantitative inaccuracies for wave height and speed for one- and two-equation models. Noting that the primary instability for the falling film is 2D, Albert et al. \cite{Albert2014} also used 2D DNS to identify the mass transfer mechanisms as vortices and backflows that occur within the main hump and first capillary ripple of a solitary wave, investigating a regime spanning $15 \leq Re \leq 53$. The onset of these vortices and flow reversals was categorised by Rohlfs and Schied \cite{Rohlfs2015} who created a phase diagram that extended from the drag-gravity to drag-intertia regimes for $\gamma_2$ type waves that are most likely to contain these flow features. To capture this phase space and maintain computational efficiency, the low-dimensional model of weighted-residuals integral boundary layer approach (WIBL) by Ruyer-Quil and Manneville \cite{Ruyer-Quil1998,Ruyer-Quil2000} was implemented and validated against a smaller sub-set of 2D DNS cases where $Re \leq 15$. Using particle image/tracking velocimetry and planar laser-induced fluorescence techniques alongside 2D DNS and the WIBL approach of \cite{Ruyer-Quil1998,Ruyer-Quil2000}, Denner et al. \cite{Denner2018} also explored solitary waves for similar inertia-dominated regimes as Albert et al. \cite{Albert2014} with $Re > 20$. This work revealed a stagnation and reduction in the maximum film height and flow rate when $Re > 60$. Considering the connection between both parameters, the reduced inertia associated with the reduction in maximum flow rate was suggested as the reason for the saturation of film height. This reduction in inertia arose due to the presence of the large flow circulation inside the main wave hump, highlighting the potential for a limiting criteria where these flow features benefit heat and mass transport.      

More recently, fully-resolved 3D DNS were used to investigate the formation of 3D horseshoe waves in falling films in the presence of surfactants \cite{Batchvarov2021}. Similar to the 2D $\gamma_2$ type waves discussed previously, wave circulations in the main wave hump and flow-reversal within the capillary ripple region were observed. However, the presence of surfactants alters the wave shape in a stabilising manner and completely suppresses these internal circulations at the highest Marangoni number tested. Other stabilising effects have been observed when introducing rotation. In the case of falling films flowing down the inner surface of a rotating cylinder, Farooq et al. \cite{Farooq2020} used 2D and 3D DNS to investigate the effects of Ekman number on wave formation, wave shape, and internal flow structures for $0 \leq Ek \leq 484$. Increasing $Ek$ was found to suppress wave formation, wave circulations, and create distinctive angled waves through the combined effect of gravity and the centrifugal force acting orthogonally to the cylinder surface. Considering a sufficiently large cylinder curvature \cite{Chen2004}, and using the analogy between this configuration and thin film flow down an inclined flat plate (where gravity is resolved  into normal and streamwise components), a second-order WIBL method \cite{Rohlfs2018} was used to create a phase map that identified the $Re-Ek$ conditions where wave circulations were either present or suppressed.     

\begin{figure*}
        \includegraphics[width=14cm]{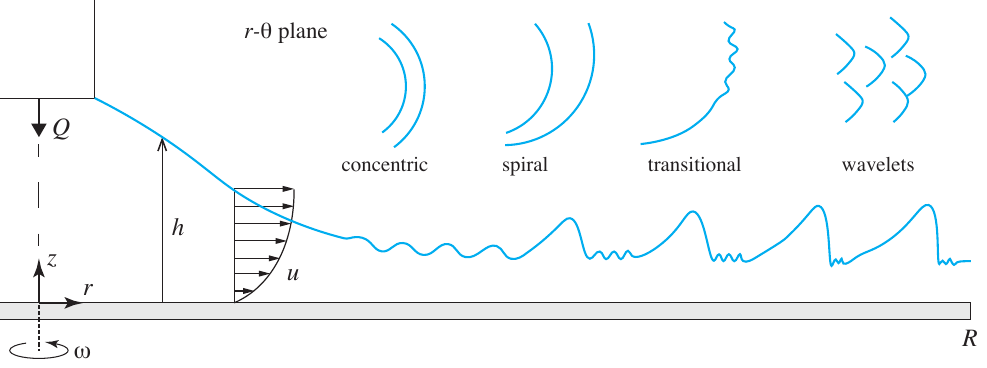}
        \caption{\label{fig:schematic}Schematic illustration of the main hydrodynamic features of a liquid film flowing over a spinning disc.}
    \end{figure*}

For thin films flowing over spinning discs, increasing rotational speed has a destabilising effect as the centrifugal force along the radial direction scales according to $\sim\omega^2 r$, where $r$ is the radial coordinate and $\omega$ is the rotational speed. The inlet flow rate ($Q$) also plays a role in the film dynamics, influencing the local film Reynolds number ($Re = Q/2\pi{r}\nu$ where $\nu$ is the kinematic viscosity) as in the falling film case. Adjusting only these two parameters ($\omega,Q$) can create a variety of wave patterns and film dynamics along a short radial length, making flows over spinning discs an attractive system for compact, continuous flow applications \cite{Matar2006}. Taking inspiration from Butuzov and Pukhovoi \cite{Butuzov1976}, Figure  \ref{fig:schematic} illustrates the range of hydrodynamic features that have been observed for film flow over a spinning disc. 
Although there are common features between falling films and films over spinning discs, there are also distinctive differences that emerge. Unlike falling films, whose dimensionless parameters are constant along the substrate length, thin films over spinning discs are influenced by several parameters that increase ($We \sim r^{4/3}$, $Ek\sim r^{4/3}$, where $We$ denotes the Weber number) and decrease with radial position ($Re\sim r^{-1}$). Experiments by Charwat et al. \cite{Charwat1972}, Azuma and Nunobe \cite{Azuma1989}, Woods \cite{Woods1995}, and Leneweit et al. \cite{Leneweit1999} reveal concentric waves and spiral waves that are stationary and unwind in the direction of rotation. The occurrence of these waves may also be sensitive to the nozzle inflow conditions \cite{Leneweit1999}. 

In recent experiments by Wang et al. \cite{Wang2023}, it was hypothesised that these spiral waves emerge from concentric waves upstream through several possible disturbances including flow rate fluctuations. However, the earlier work by Charwat et al. \cite{Charwat1972}, which combined experimental observations with a linear stability analysis, suggest that spiral waves form at the point when Coriolis forces become significant, $Ek \sim 1$. Here, they found that when $0.5 < Ek < 2.5$, the Coriolis force may have a stabilising effect on the imposed concentric waves and the most likely scenario for spiral wave development is through the indirect effect of creating a non-unidirectional mean flow. Laser doppler velocimentry measurements have indicated the growth of a fully-3D laminar velocity profile within the film as the fluid moves downstream on the disc \cite{Azuma1989}. In falling films, 2D solitary waves transition into localised 3D horseshoe or so-called `scallop' waves that have few coalescence events and maintain their density \cite{Chang2002,Batchvarov2021}. In contrast, and as we will show in this work, wavelet coalescence is a prevalent characteristic in the case of thin films flowing over a spinning disc. The importance and interdependence of these aspects have provided challenges in studying the dynamics of these films in detail through high-fidelity methods, and this work aims to address this shortcoming.

Previous simulation studies of thin film flows over a spinning disc have implemented low-dimensional methods \cite{Sisoev2003,Matar2004,Matar2005,Kim2024} and under-resolved computational fluid dynamics approaches such as LES and RANS \cite{Li2020}. The former are based on various adaptations of the integral boundary layer approach. Sisoev et al. \cite{Sisoev2003} modelled the formation of waves on thin films under axisymmetric assumptions and in the large Ekman number limit, where solutions to the falling film problem were found to provide a reasonable description. Using the Karman-Polhausen approach and an assumption of a semi-parabolic velocity profile within the film, Matar et al. \cite{Matar2004} derived and solved evolution equations for film height and flow rate in the radial and azimuthal co-ordinate space. This revealed the spatio-temporal development of $\gamma_1$ and $\gamma_2$-type waves with large-amplitudes along the disc radius, while also assessing their impact on mass transfer behaviour \cite{Matar2005}. Two governing dimensionless parameters, the dimensionless wavelength, $\lambda$ (similar to an inverse Weber number), and a modified Ekman number, $r_{\text{disc}}$, were also derived. These condense several traditional parameters (e.g., $Re$, $We$, $Ca$, and $Ek$) into two that concisely describe the hydrodynamics of thin film flows on spinning discs. 

Recently, an extended integral boundary layer method for 3D surface waves was proposed which also captures turbulent flow within the film that may occur due to the presence of internal vortex structures near the nozzle, as well as laminar flow that exists downstream due to film thinning \cite{Kim2024}. Closure models for each condition implemented either turbulent (power-law) or laminar (polynomial) boundary layer approximations that were selected by assessing a local Reynolds number based on the horizontal depth-averaged velocity (the transition to turbulence was set at $Re_L = h|\mathbf{\overline{u}}|/\nu > 100$, wherein $\overline{u}$ is the average film velocity). The effects of `backscatter', a stochastic model introduced to represent the transfer of energy within the film from small 3D eddies to large 2D eddies, were assessed by varying an empirical constant within the model and showed a significant impact on surface wave formation along the disc.  
Despite these previous efforts, insights into the film dynamics in low-$\lambda$ regimes where inertia plays an important role ($0.001 \lesssim \lambda \lesssim 0.01$) are limited, with most studies focusing on flows where inertial effects are less prevalent ($\lambda > 0.01$) and using low-order methods to reconcile observations with experiments. While reduced order models have provided important insights in this regard, they also preclude a rigorous analysis on the film dynamics in these nonlinear regimes by introducing assumptions about the nature of the flow within the film. 

In this work, we examine the film dynamics and evolution of surface waves using direct numerical simulations in combination with high-speed imaging experiments. The paper first introduces the experimental setup in Section \ref{sec:Exp-methods} along with the numerical method and the validation in Section \ref{sec:DNS-method}. This is followed by a discussion of our results from this investigation in Section \ref{sec:Results}. We use the $\lambda$ and $r_{disc}$ scaling proposed by Matar et al. \cite{Matar2005} to classify the wave types and construct a phase diagram that covers a space that is both diverse in wave characteristics and operationally important for practical applications of spinning disc processors ($0.001 < \lambda < 0.035$ and $1 < r_{disc} < 10$). Upon constructing this phase diagram, we then probe the effects of different wave characteristics on strain rates both on the disc surface and within the film depth. We explore the wave development all the way from a smooth or waveless interface to 3D wavelets, revealing the transitional elements and internal flow mechanisms.   

\section{\label{sec:Exp-methods}Experimental Method}

\begin{figure*}
        \includegraphics[width=14cm]{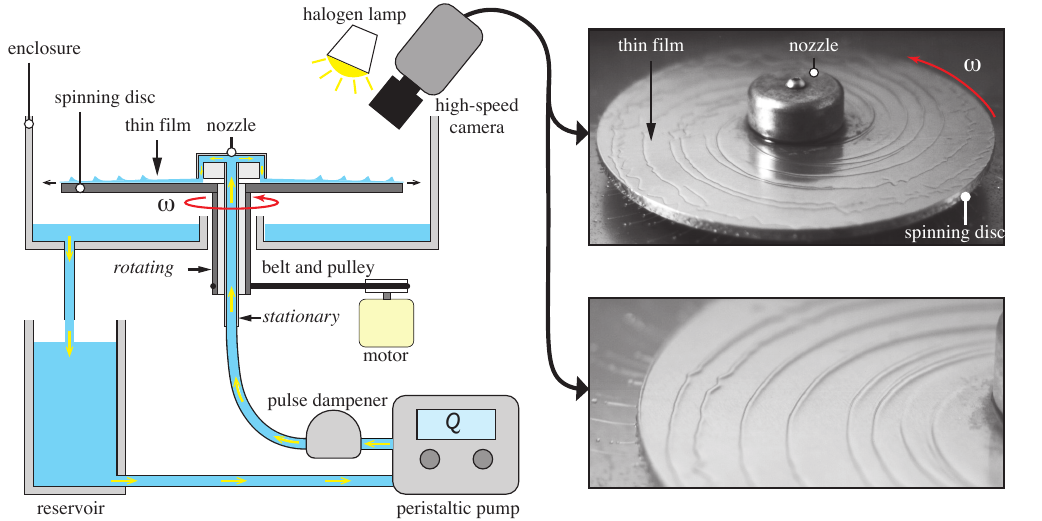}
        \caption{\label{fig:1}Experimental setup for generating and imaging thin liquid films over a rapidly rotating disc. For a Newtonian fluid with constant properties ($\rho_l,\mu_l,\sigma$), the pump flow rate ($Q$) and disc rotational speed ($\omega$) control the hydrodynamic behaviour.}
    \end{figure*}

\noindent A smooth rotating disc with a diameter of 20 cm was used in this study and manufactured from stainless steel. This was encased in a static enclosure housing consisting of a 304 stainless steel box, 45 cm in width (Fig. \ref{fig:1}). The top of the enclosure was left open to the ambient air to allow for flow visualisation experiments. The rotational speed of the disc ($\omega$) was controlled using an Oriental Motor GmbH, 2016 model: BMU260C-AC-3 and belt and pulley system as shown in Fig. \ref{fig:1}. This was capable of delivering rotational speeds up to 4000 rpm and allowed fine tuning of the speed down to 1 rpm increments. Rotational speeds of up to 2000 rpm were investigated as this produced a sufficiently wide range of thin film flow regimes considered for the current work. 

To control the inlet flow rate ($Q$) of the fluid, a peristaltic pump was employed to circulate liquid from a reservoir, through a stationary nozzle, and onto the rotating disc. The enclosure housing contained an outlet port to return the liquid back to the reservoir using gravity. The peristaltic pump used was a Masterflex 07554-95 (Cole-Parmer), and capable of delivering flow rates in the range 0.02--48 mL.s$^{-1}$. To remove any effect of pulsation from the pump on the hydrodynamics on the disc, a pump dampener was employed (Masterflex 7596-20). Flow rates in the range 6--26 mL.s$^{-1}$ were selected for the evaluation of the hydrodynamics on the disc. N-methyl-2-pyrrolidone (NMP) was used as the working liquid for the experiments, a common solvent in chemical manufacturing processes. At room temperature (25$^{\circ}$C), NMP has a surface tension ($\sigma$) of 40.7 mN.m$^{-1}$, density ($\rho_l$) of 1027 kg.m$^{-3}$, and dynamic viscosity ($\mu_l$) of 1.65 cP. The liquid was introduced from the bottom of the disc to allow for a more compact design, and the injection was facilitated by means of the stationary nozzle. The diameter of the nozzle housing was 46 mm, with the liquid exiting onto the disc through an annular slot jet of 1 mm width and in close proximity to the target surface. The choice of this type of nozzle design over a  circular jet was made to avoid introducing unsteadiness in the film entrance region which can emerge from shear-driven disturbances at the interface of a circular jet prior to impingement on the disc surface \cite{Kim2024}.  

The thin film wave regimes over the spinning disc were experimentally observed using high-speed imaging with an Olympus i-Speed 3 camera and a 60 mm or 105 mm Nikon AF-S Nikkor (f/2.8) lens. The camera frame rate was selected based on the spinning disc operational conditions and covered a range from 4000--10000 fps. Exposure time was adjusted to be in the range 30--100 $\mu$s to balance image intensity and avoid blurring/streaking of the interfacial waves. Lens magnification and aperture settings were adjusted for full disc and zoomed disc images of the thin films forming over the disc surface (see examples in Fig. \ref{fig:1}). A halogen lamp (400W) with a light diffuser sheet was used to illuminate the region of interest. Heating of the liquid film and experimental apparatus was limited by only operating the lamp during the image acquisition stage which typically lasted $< 60$ s. The entire apparatus, excluding camera, was placed inside a laboratory fumehood for safety and consistent environmental conditions. Liquid and ambient air temperatures were monitored using Type-K thermocouples and this arrangement was found to provide a constant temperature during testing.   

\section{\label{sec:DNS-method}Problem Formulation \& Numerical Methodology}

\subsection{\label{subsec:problemFormulationDNS}Governing equations \& waveless flow}

\noindent In the spinning disc arrangement, a liquid of density $\rho_l$ and viscosity $\mu_l$ is introduced at a flow rate $Q$ from a central nozzle and distributed as a thin film on a rigid and impermeable disc of radius $R$, which rotates at a constant rotational speed $\omega$. The liquid film is in contact with a gas phase of density $\rho_g$ and viscosity $\mu_g$ through a gas-liquid interface with surface tension $\sigma$. The main forces at play that influence the thin film dynamics are the centrifugal force, viscous shear, and surface tension. We use  cylindrical co-ordinates ($r$, $\theta$, $z$), defined as shown in Fig. \ref{fig:2} for a disc sector of angle $\phi$, with associated velocity components in the radial, azimuthal, and axial directions ($u_r$, $u_\theta$, $u_z$, respectively) to describe the flow. The fluid is considered incompressible, Newtonian, and isothermal. Accordingly, the thin film dynamics can be described by the continuity and Navier-Stokes equations:
\begin{eqnarray}
    \label{eq:continuityNS}
    \nabla \cdot \bf u = &&\,0,\\
    \label{eq:momentumNS}
    \rho \left(\pderiv{u}{t} + \bf u \cdot \bf \nabla u \right) =&&\,-\nabla p +\mu \nabla^{2} \mathbf{u} + \rho g{\bf\hat{z}} + \sigma{\kappa}\mathbf{n}{\delta},
    \label{eq:NS}
\end{eqnarray}
\noindent where $\bf u$ is the velocity, $p$ the pressure, $g$ the gravitational acceleration, $\kappa$ the curvature of the interface, $\mathbf{n}$ is the unit normal at the interface, $\delta$ (not to be confused with the reduced Reynolds number parameter used in falling film theory mentioned in Sec. I) represents the Dirac delta function concentrated at the interface, and $\mathbf{\hat{z}}$ is the unit vector in the $z$ direction. 
Taking the momentum equation, Eq. (\ref{eq:momentumNS}), and examining the flow in the rotating reference frame yields
\begin{equation}
    \label{eq:momentumFoR}
    \rho\left(\frac{\partial \mathbf{u}^{\prime}}{\partial t}+\mathbf{u}^{\prime} \cdot \nabla \mathbf{u}^{\prime}\right)=-\nabla p+\mu \nabla^2 \mathbf{u}^{\prime}+\rho \mathbf{g}+\sigma \kappa \mathbf{n} \delta-\boldsymbol{\omega} \times(\boldsymbol{\omega} \times r)-2 \boldsymbol{\omega} \times \mathbf{u}^{\prime},
\end{equation}
\noindent where $\mathbf{u}^{\prime} = \mathbf{u} - \omega{\mathbf{r}}$, 
and $\boldsymbol{\omega} = \omega{\mathbf{\hat{z}}}$. From this re-framing, the centrifugal and Coriolis forces emerge and are represented by the last two terms on the right-hand side of Eq. (\ref{eq:momentumFoR}). 

We use a single-fluid, volume-of-fluid (VoF) \cite{Hirt1981}, interface-capturing approach to model the interfacial dynamics whereby the surface tension force at the interface separating the liquid and gas phases, corresponding to the $\sigma \kappa \mathbf{n}\delta$ term in  Eq. (\ref{eq:NS}),
is modelled according to the Continuous Surface Force (CSF) method \cite{Brackbill1992}. Here, $\mathbf{n}$ and $\kappa$ are respectively expressed as:
\begin{equation}
    \mathbf{n}=\frac{\nabla\alpha}{|\nabla\alpha|}, ~~~
    \kappa=-\nabla\cdot \left(\frac{\nabla\alpha}{|\nabla\alpha|}\right),
\end{equation}
\noindent and $\alpha$ is the volume fraction of the liquid phase, varying from 0 in the gas to unity in the liquid. 
The local density $\rho$ and viscosity $\mu$ are given by the 
following $\alpha$-weighted expressions:
\begin{eqnarray}
    \label{eq:rho_avg}
    \rho = \alpha \rho_l + (1 - \alpha)\rho_g, \\
    \label{eq:mu_avg}
    \mu  = \alpha \mu_l + (1 - \alpha)\mu_g. 
\end{eqnarray}
A species transport equation is solved to describe the advection of the liquid-phase volume fraction $\alpha$, with the form:
\begin{equation}
    \label{eq:alphaAdv}
    \pderiv{\alpha}{t} + \nabla \cdot (\alpha {\bf u}) = 0. 
\end{equation}
This equation is not solved directly in the VoF method implemented in this work. The interface location is first
identified and then advected in a geometric manner as described below in Sec. \ref{subsec:numericalDNS}. 

The governing equations admit a steady, waveless solution, which is equivalent to the Nusselt solution obtained for falling films, which reflects a balance between centrifugation and viscous drag. Applying lubrication theory \cite{Emslie1958} and neglecting advective, gravitational ($g \ll \omega^{2}r$), surface tension, and Coriolis terms in Eq. (\ref{eq:momentumFoR}) yields:
\begin{equation}
    \label{eq:simMom}
    -\mu_l \frac{\partial^2 u}{\partial z^2}=\rho_l\omega^2 r,
\end{equation}
where $u$ is the radial velocity component. Integrating and applying no-slip boundary conditions at the disc surface ($u = 0$ at $z = 0$) and zero-shear at the interface ($\partial u / \partial z = 0$ at $z = h$) as illustrated in Fig. \ref{fig:schematic}, results in an expression for $u$:
\begin{equation}
    \label{eq:uNlocal}
    u=\frac{\omega^2 r}{2 \nu_l}\left(2 h z-z^2\right),
\end{equation}
\noindent where $h$ is the height of the interface. The average velocity across the film thickness, $u_N$, is given by:
\begin{equation}
    \label{eq:uN}
    u_{N}=\frac{1}{h_N} \int_0^{h_N} u \mathrm{~d} z=\frac{\omega^2 r h^2_N}{3 \nu_l},
\end{equation}
\noindent where $h_N$ is the waveless, Nusselt film thickness and
$\nu_l$ is the kinematic viscosity of the liquid. 
The analogy between the spinning disc and the falling film arrangement for vertical drainage is observed when replacing $\omega^2 r$ by the gravitational acceleration, $g$, in Eq. (\ref{eq:uN})  \cite{Aoune1999}.
For a constant volumetric flow rate $Q$ introduced from the nozzle (see Fig. \ref{fig:schematic}), we have the following expression:
\begin{equation}
    Q = 2 \pi r h_N u_{N}=\frac{2 \pi r^2 \omega^2}{3 \nu_l} h^3_N.
\end{equation}
\noindent Rearranging gives the following expression, $h_N$:
\begin{equation}
    \label{eq:hN}
    h_N=\left(\frac{3 \nu_l Q}{2 \pi r^2 \omega^2}\right)^{1 / 3},
\end{equation}
\noindent which in the following section we will use to initialise the direct numerical simulations. Note that unlike the falling film case in which $h_N$ is a constant, in the case of flow over a spinning disc, $h_N \sim r^{-2/3}$. 
The wall shear rate for waveless flow conditions is obtained by differentiating Eq. (\ref{eq:uNlocal}) and evaluating the result at the disc surface:
\begin{equation}
\label{eq:gammaNlocal}
    \dot{\gamma}_{w,N} = \frac{d u}{d z}\bigg\vert_{z=0} = \left(1.5\frac{r \omega^4 Q}{\nu_l^2\pi}\right)^{1/3}.
\end{equation}

We can also define common dimensionless parameters that describe the film hydrodynamics:
\begin{equation}
    Re = \frac{h_N u_N}{\nu_l} =  \frac{Q}{2\pi r \nu_l},
\end{equation}
\begin{equation}
    We =\frac{\rho_l h_N u_N^2}{\sigma} =  \frac{\rho_l}{\sigma}\left(\frac{Q}{2 \pi}\right)^{5 / 3}\left(\frac{\omega^2}{3 r^4 \nu_l}\right)^{1 / 3},
\end{equation}
\noindent where $Re$ and $We$ are the Reynolds and Weber numbers, respectively, based on $h_N$ nd $u_N$ as the characteristic film height, Eq. (\ref{eq:hN}), and velocity, Eq. (\ref{eq:uN}), for waveless flow, respectively. In addition, and specific to the spinning disc arrangement, the Ekman number describes the significance of viscous to Coriolis effects:
\begin{equation}
    \label{eq:Ek}
    Ek =\frac{\nu_l}{\omega h_N^2} = \left(\frac{2\pi}{3Q}\right)^{2/3}\left(\omega r^4 \nu_l\right)^{1/3}.
\end{equation}
\noindent As noted in Sec. \ref{sec:Introduction}, the relationships above highlight the dependence of the dimensionless groups on the radial coordinate $r$, unlike the case of falling films wherein they are constant. 

In this study, we reduce the number of dimensionless parameters from three to two by adopting the scaling relations of Matar et al. \cite{Matar2005} who identified the dimensionless wavelength $\lambda$, similar to an inverse $We$, and a modified Ekman number $r_{\text{disc}}$ to describe the film hydrodynamics for a spinning disc, respectively expressed by
\begin{equation}
    \label{eq:lambda}
    \lambda^2 = \frac{\sigma}{\rho_l} \left(\frac{2\pi}{Q} \right)^2 \left( \frac{\nu_l}{\omega} \right )^{3/2}, 
\end{equation}
\begin{equation}
    \label{eq:rdisc}
    r_{\text{disc}} = R \left(  \frac{2 \pi}{Q} \right )^{1/2} (\nu_l \omega)^{1/4}.
\end{equation}
The connection between $r_{\text{disc}}$ and the Ekman number is evident when considering Eq. (\ref{eq:Ek}) whence  $r_{\text{disc}} \equiv Ek^{3/4}$. Using these scaling relations, the interplay of inertia, viscosity, surface tension, and the Coriolis force is captured by the dimensionless wavelength $\lambda$ and the modified Ekman number $r_{\text{disc}}$, which effectively govern the emergence of the various flow regimes observable on the disc. As will be shown in the following sections, the formation of interfacial waves on a thin film flowing over a spinning disc is favoured by low $\lambda$ and large $r_{\text{disc}}$ values. 

\subsection{\label{subsec:numericalDNS}Numerical method}

\noindent The governing equations for continuity, momentum and species transport, corresponding to Eqs. (\ref{eq:continuityNS}), (\ref{eq:momentumNS}), and (\ref{eq:alphaAdv}) were solved using the finite-volume method implemented within OpenFOAM. The coupling of velocity and pressure is performed using the Pressure-Implicit with Splitting of Operators (PISO) algorithm. In order to construct a sharp interface, %
Eq. (\ref{eq:alphaAdv}) is modified to compress the surface and reduce smearing \cite{Deshpande2013}. In this work, we implement the geometric isoAdvector method developed by Roenby {\it et al.} \cite{Roenby2016}, which uses the concept of isosurfaces to calculate more accurate face fluxes, specifically for the cells containing the interface, allowing for higher Courant numbers ($Co \approx 0.5)$ than algebraic methods such as MULES. Here, the isosurface of the gas-liquid interface is located at the locus of points for which $\alpha = 0.5$. 

    \begin{figure*}
        \includegraphics[width=14cm]{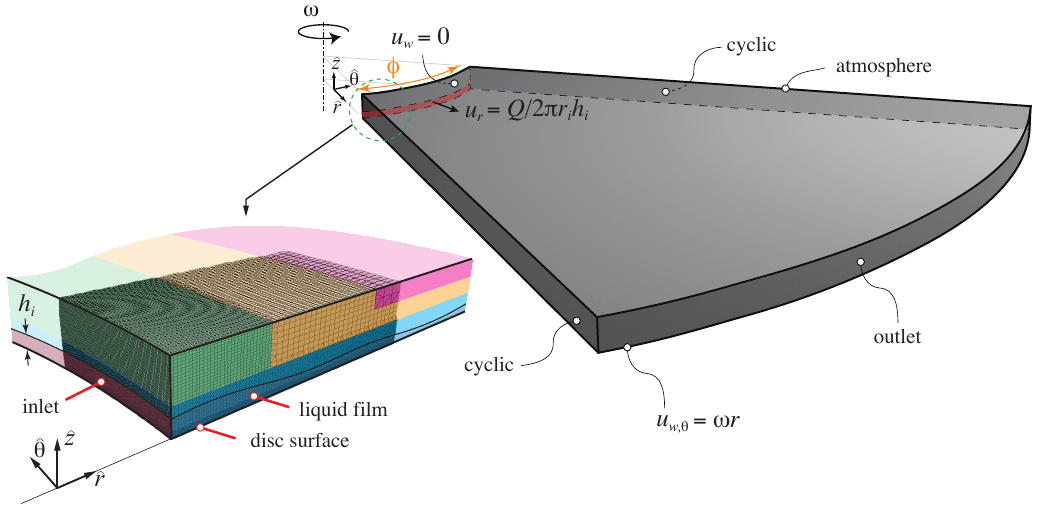}
        \caption{\label{fig:2}Computational domain for the direct numerical simulation (DNS) of thin film flows over a spinning disc (right). The grid (left) was split into multiple graded zones to resolve the interfacial thin film flow (blue zone) and far field gas flow (green, yellow, and pink zones).}
    \end{figure*}

To reduce the resource demands for DNS, the computational domain comprised a sector of the disc (Fig. \ref{fig:2}). 
In this arrangement, periodic, or cyclic, conditions are used to map properties of the flow leaving one boundary onto the corresponding boundary where flow enters. The concept proposed is that assuming sufficiently large sector angles ($\phi$) and simulation time, the flow within the domain will be representative of a complete spinning disc. A suitable sector angle was determined based on comparisons between experiments and predictions, and varied in the range of $25^{\circ} \leq \phi \leq 60^{\circ}$ across the cases examined in this study. The other boundary conditions include a no-slip moving wall ($u_{w,\theta} = \omega r$) for the disc surface, uniform radial velocity at the nozzle inlet ($u_{r} = Q/2\pi{r_i}h_i$, where $h_i$ and $r_i$ are the nozzle inlet height and radius, respectively), and a no-slip boundary for the nozzle section above the inlet ($u_w = 0$). To represent the flow leaving the domain at the outlet (or gas flow entering the domain from the atmosphere above the disc), a zero-gradient condition was set for positive fluxes out of the domain (or negative fluxes into the domain).  

The numerical grid was constructed by first extruding a single $r-z$ plane ($L_r \times L_z = [R - r_{i}] \times 4h_i$) of hexahedral cells in the $\theta$ direction across the required sector angle. The nozzle inlet height, radius, and disc radius, were chosen to match the experimental values (e.g., $h_i = 1$mm, $r_i = 23$mm, $R = 0.1$m). Subsequent refinements were performed on the computational grid to increase the density of cells in the thin film region, as shown in Fig. \ref{fig:2}. This allowed for resolving the thin film flow phenomena while also reducing the computational demands in the far field gas flow regions. The final grid size for the cases examined in this work was up to a maximum of 70 million cells. With maximum Courant number constraints in place, the numerical schemes imposed were backwards Euler for time derivatives and Gauss linear for gradient and divergent terms.

The properties of the liquid phase were taken from the experimental conditions and described in Sec. \ref{sec:Exp-methods}. For the gas phase, density $\rho_g = {1.18}$ kg.m$^{-3}$ and kinematic viscosity $\nu_g = 1.57\times10^{-5}$ m$^2$.s$^{-1}$ were used for air.
\noindent The DNS were validated against experiments performed in this study and using the experimental findings from Woods \cite{Woods1995}. For the latter, a different spinning disc setup was modelled, with a disc radius of 0.18 m, nozzle radius of 41 mm, and nozzle height of 0.5 mm. The liquid phase was a mix of water and nigrosene dye with properties $\rho_l = 1000$ kg.m$^3$, $\mu_l = 1.004$ cP, and $\sigma = 49.5$ mN.m$^{-1}$, for the liquid density, viscosity, and surface tension, respectively. 

\begin{figure*}
        \includegraphics[width=14cm]{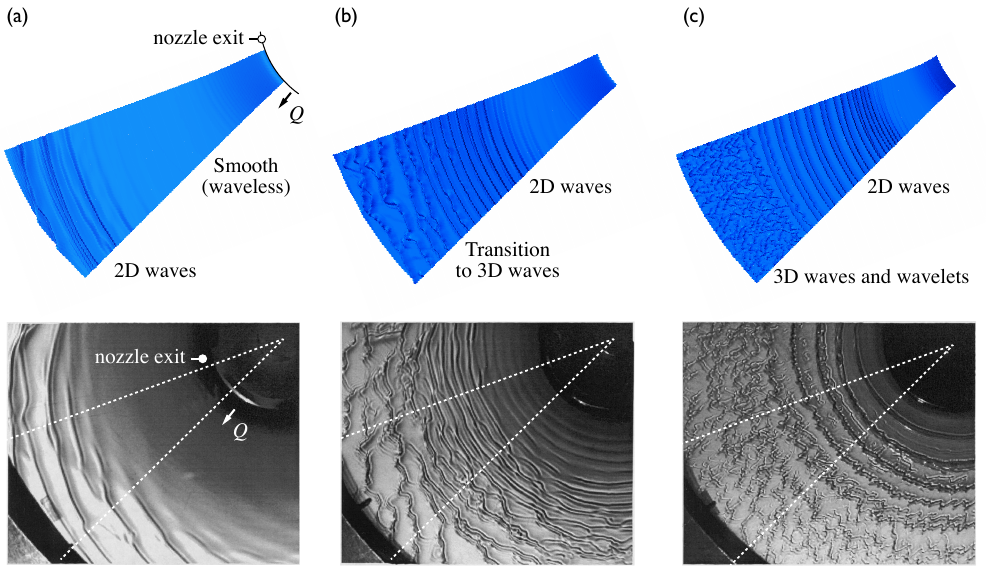}
        \caption{\label{fig:3}Comparison between the liquid-gas interface predicted using DNS (top) and the experiments of Woods \cite{Woods1995} for (a) $\lambda = 0.013$, $r_{disc} = 5.95$, (b) $\lambda = 0.0077$, $r_{disc} = 7.07$, and (c) $\lambda = 0.0034$, $r_{disc} = 9.31$.}
    \end{figure*}

\begin{figure*}
        \includegraphics[width=14cm]{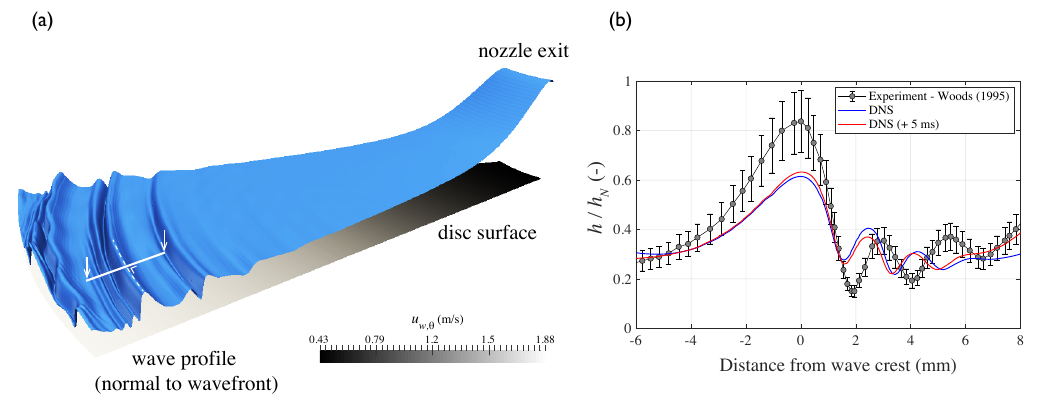}
        \caption{\label{fig:4}Analysis of the wave profile height predicted using DNS (a) and compared with our experimental measurements taken normal to the wavefront at the same radial location (b) for $\lambda = 0.013$, $r_{disc} = 5.95$.}
    \end{figure*}

A section of the gas-liquid interface predictions is compared with the experiments by Woods \cite{Woods1995} at three different combinations of $\lambda$ and $r_{disc}$ values in Fig. \ref{fig:3}. These test cases cover smooth or waveless flow, 2D waves, transition-to-3D waves, and 3D waves, providing a comprehensive set of hydrodynamic conditions to compare the model against. The predictions were found to accurately capture both the change in wave regime and the structure of the waves observed in the experiments. 
In Fig. \ref{fig:4}, film height and wave shape of a 2D spiral wave predicted by the DNS are quantitatively compared with the experimental measurements obtained by Woods \cite{Woods1995} for the same wave regime ($\lambda = 0.013$). The wave shape, comprising the main wave hump and preceding capillary ripples, are closely predicted with local differences in film thickness of 20--30\%. We believe this to be acceptable as slight spatial and temporal variations are expected given the transient and stochastic nature of the investigated flow, and when accounting for the uncertainty in the measurements of film thickness within the experiment. 

\section{Results and Discussion}\label{sec:Results}

\noindent We begin our discussion with an analysis of the wave regimes that emerge for flow over a spinning disc. Following this, we explore film thickness statistics and the impact this and the wave regimes have on the wall shear rate distributions. The evolution of wave types is then examined to reveal the mechanisms and internal dynamics of thin films flowing over spinning discs.

\subsection{\label{sec:Flow-regimes}Flow regimes and phase diagram}
\begin{figure*}
        \includegraphics[width=14cm]{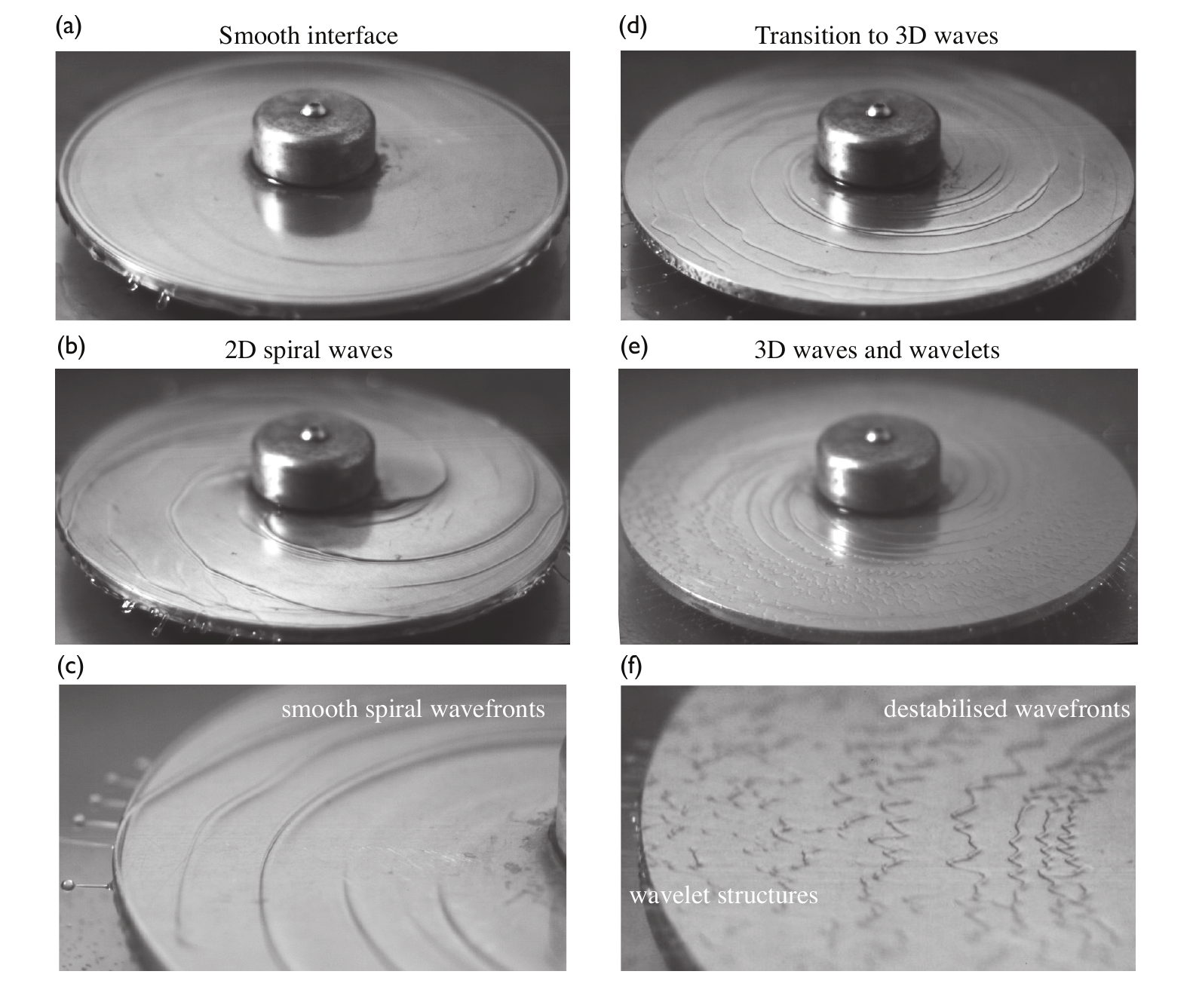}
        \caption{\label{fig:5}Classification of the interfacial flow regimes from a smooth interface to the occurrence of 3D waves (a-e). The transition from smooth spiral wavefronts (c) to 3D waves (f) begins with a destabilisation of the wavefronts followed by breakup into wavelets at larger radial distances downstream. The experimental parameters are (a) $\lambda = 0.052$, $r_{disc} = 6.57$, (b) $\lambda = 0.0116$, $r_{disc} = 4.78$, (c) $\lambda = 0.0127$, $r_{disc} = 5.80$, (d) $\lambda = 0.0184$, $r_{disc} = 9.29$, (e) $\lambda = 0.0035$, $r_{disc} = 7.15$, and (f) $\lambda = 0.0093$, $r_{disc} = 11.68$.}
    \end{figure*}

\noindent Through examination of the high-speed images, DNS predictions, and comparison with previous studies in the literature (e.g. Fig. \ref{fig:3}), common interfacial flow features were revealed, which can be classified into distinct regimes over the spinning disc. Examples of these regimes are shown in Fig. \ref{fig:5}. Increasing inertial effects by reducing $\lambda$ and/or increasing $r_{disc}$ leads to a progression from initially smooth or waveless flow (a), to the formation of 2D spiral waves (b,c), the destabilisation of spiral wavefronts during the transition to 3D waves (d,f), and finally, complete spiral wave breakup into wavelet flow structures (e,f). We observed these 3D wavelets travelling in isolation ($\Lambda$ solitons \cite{Demekhin2010}), and also colliding and merging with other waves over the disc.

\begin{figure*}
        \includegraphics[width=14cm]{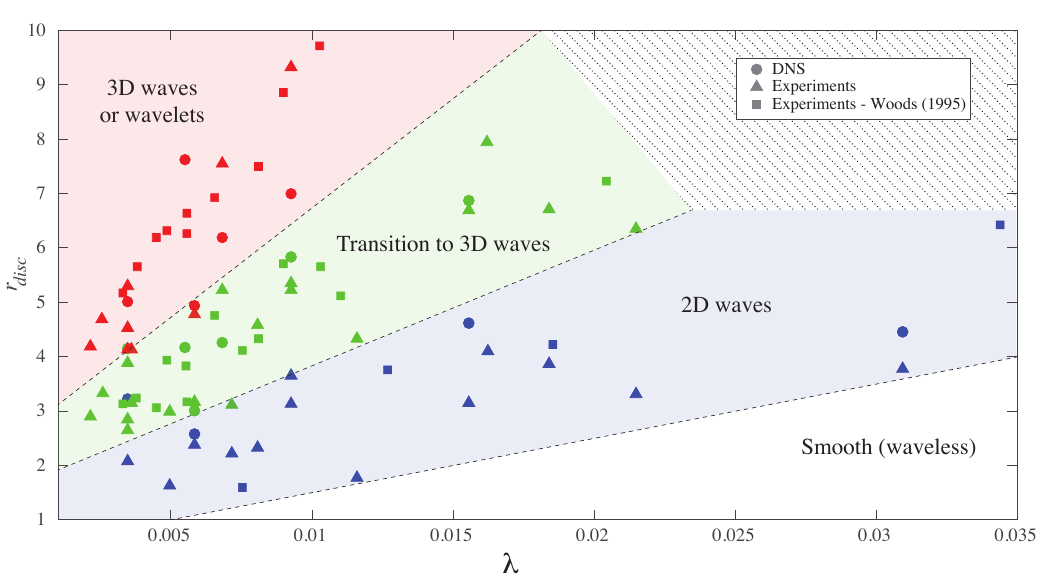}
        \caption{\label{fig:7}Phase diagram illustrating the interfacial wave regimes across the $\lambda-r_{disc}$ space.}
    \end{figure*}

By mapping out the different regimes observed in our experiments and numerical predictions, we constructed a phase diagram to understand the relationship between the interface features and the main parameters used in this work to describe the hydrodynamics of spinning disc flows (Fig. \ref{fig:7}). In many cases, multiple flow regimes exist on the spinning disc (Fig. \ref{fig:5}). We have previously defined $\lambda$ and $r_{disc}$ as global parameters, given by Eqs. (\ref{eq:lambda}) and (\ref{eq:rdisc}), respectively, which describe the hydrodynamics. To construct the map, we use the \textit{local} radius as the characteristic length and replace the disc radius in Eq. 
(\ref{eq:rdisc}) to identify the boundaries between different flow regimes on the disc. This way, the phase diagram can be interpreted using $r_{disc}$ based on either the local or disc radius as characteristic length scales. For example, using the disc radius to calculate $r_{disc}$ in Eq. (\ref{eq:rdisc}), a horizontal line can be plotted through the values of $\lambda$ on the abscissa of Fig. \ref{fig:7} to $r_{disc}$, and the fraction of the disc that will be covered by the various wave regimes can be obtained.   

Interestingly, within the \textit{2D waves} regime of the phase map, concentric waves were not observed in either the experiments or DNS results. This confirms that concentric waves are not required to form spiral waves, contrary to recent suggestions \cite{Wang2023}. Instead, we find that it is the Coriolis effect, which underlies the spiral wave formation that unwind in the direction of rotation, similar to the observations from earlier works investigating thin films over spinning discs \cite{Charwat1972,Leneweit1999}. This suggests that differences in nozzle configuration govern the likelihood of concentric waves emerging. In our work, the nozzle design shown in Fig. \ref{fig:1} provides a steady inflow to the central region of the disc. Other studies utilise an impinging jet, which can form an unstable free surface prior to reaching the target that may introduce unsteadiness that propagates into the film layer to form concentric waves near the inlet region (e.g. \cite{Kim2024}). 

The temporal behaviour of the waves was also found to be dependent on the wave regime. The spiral waves that form are stationary in the laboratory reference frame and this can be seen in the supplementary information (video) showing both the experimental and DNS cases. Crossing the boundary between the \textit{2D wave} and \textit{transition} regimes, the spiral wavefront destabilises, as shown in Fig. \ref{fig:5}d,f, shifting into an unsteady behaviour that continues through the wavelet formation process. In the next sections, we examine the impact of these destabilisation and wave breakaway processes on the internal film dynamics. Importantly, this phase diagram provides a simple criterion for assessing reduced order models over a broad range of flow regimes, including the low-$\lambda$ conditions ($\lambda \sim 0.001$) which have received less attention. Given the simplicity of the spinning disc configuration, wherein the interfacial dynamics are governed by two operational parameters $Q$, and $\omega$, it can also serve as a reference to select and tune flow regimes for fundamental and applied investigations.

\subsection{\label{sec:WSS}Film thickness and wall shear rate distributions}
\begin{figure*}
        \includegraphics[width=14cm]{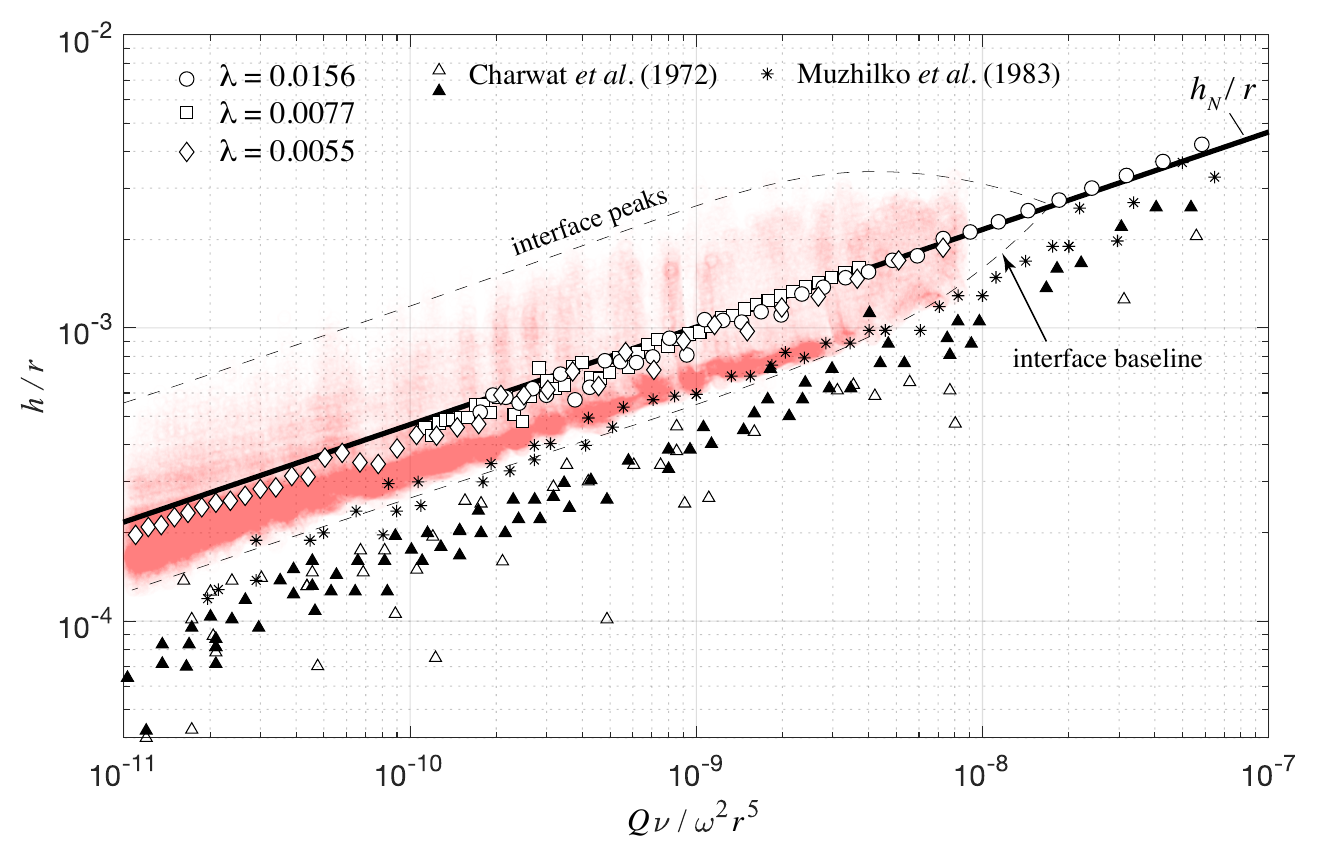}
        \caption{\label{fig:6}Average film height compared to the waveless solution (Eq. \ref{eq:hN}). The local film height across the interface surface for $\lambda = 0.0055$ are also shown (red), together with experimental data \cite{Charwat1972, Muzhilko1983} for different fluids and surfactant solutions (solid marker). }
    \end{figure*}

\noindent The film thickness is a key parameter in thin film flows owing to its influence on transport phenomena. An open question over the last several decades has been the inability to reconcile previous experimental observations of an apparent decrease in film thickness relative to the Nusselt solution for $Q\nu/\omega^2 r^5 < 10^{-7}$ shown in Fig. \ref{fig:6}, which coincides with the presence of wavy flow. Recalling the Nusselt solution for film thickness given by Eq. (\ref{eq:hN}), this parameter has been used here to represent the data as $h_N/r = (3Q\nu/2\pi\omega^2 r^5)^{1/3}$. Charwat et al. \cite{Charwat1972} investigated whether gas flow above the disc, which can be driven by the rapid rotation of the film layer \cite{Butuzov1976}, could be the reason for the differences observed. Given that exceedingly small film thicknesses can occur in this flow ($h \sim 1 - 10\mu$m), it has also been suggested that measurement error could be the main factor for discrepancies with the theory \cite{Leneweit1999}.

As we model both the liquid film and gas phases using our DNS approach, we were able to examine the hypothesis of Charwat et al. \cite{Charwat1972} in detail. Figure \ref{fig:6} shows the predicted film thickness spatially-averaged in the azimuthal direction according to $h_{avg} = (1/A)\int_s h dA$, where $S$ is a circumferential region on the disc with area $A$. We find that the averaged film thickness agrees with the Nusselt solution confirming that the dominant physical phenomena are captured by Eq. (\ref{eq:hN}) for this wide range of conditions, and that the lower film thicknesses observed in the experiments must be related to the measurement routine rather than the gas phase dynamics. 

Although the measurement error is likely to have played an important role in the work of Charwat et al. \cite{Charwat1972}, it does not fully explain  why there is reasonable agreement between independent studies on the departure of $h/r$ from theory ($10^{-8} < Q\nu/\omega^2 r^5 < 10^{-7}$), and how this departure could be exclusively attributed to measurement error in a region where the film height is at a similar scale ($h/r \sim 10^{-3}$). We examined the distribution of the local interface height and plotted this in Fig. \ref{fig:6} for the case of $\lambda = 0.0055$ (red data points). Note that we observed similar distributions for other cases and we have therefore omitted them from the figure for clarity. The density of the data shows how the film comprises a small number of tall interface peaks (waves) surrounded by large regions of the interface that are below the height of the Nusselt solution. We find that the low height, interface baseline region coincides with the trends observed in experiments \cite{Charwat1972,Muzhilko1983}. 

The bias in the experimental data towards the interface baseline mentioned above is therefore related to the film height statistics. This emphasises important considerations in the study of thin films over spinning discs: single point/region measurements of film thickness should consider both spatial and temporal variations of the interface. The presence of stationary spiral waves in the laboratory reference frame requires that average film height measurements are performed around the disc. As these spiral waves destabilise to form wavelets, they also seed the non-stationary intermittent 3D waves downstream, potentially leading to other spatial correlations in wave-induced interfacial turbulence. Considering the peak-average-baseline variations in interface height presented in Fig. \ref{fig:6}, the application of spatio-temporal statistical analyses, which are well-established for investigating turbulent flows \cite{Stafford2012}, could help inform the sampling criteria for future experiments.

The interface peaks observed in Fig. \ref{fig:6} grow to a reasonably consistent amplitude of $h_{peak}\approx2h_N$ in the regions where $\gamma_2$ and $\Lambda$ wave families exist (the interface topology will be discussed in Sec. \ref{sec:Evolution} below). In the spiral wave development stage, $\gamma_1$ waves have wave heights $h_{peak} < 1.5h_N$, approaching $h_N$ in the waveless region. Analogous to the observations for these wave types in falling films on inclined plates \cite{Liu_Paul_Gollub_1993}, lower amplitude waves travel at lower wave speeds. Considering that film height over the spinning disc scales as $h_N \sim r^{-2/3}$, reductions in the absolute wave peak height with disc radius are found. This introduces favorable conditions for wave interactions and coalescence events as the spirals unfold and destabilise with increasing disc radius. 

\begin{figure*}
        \includegraphics[width=14cm]{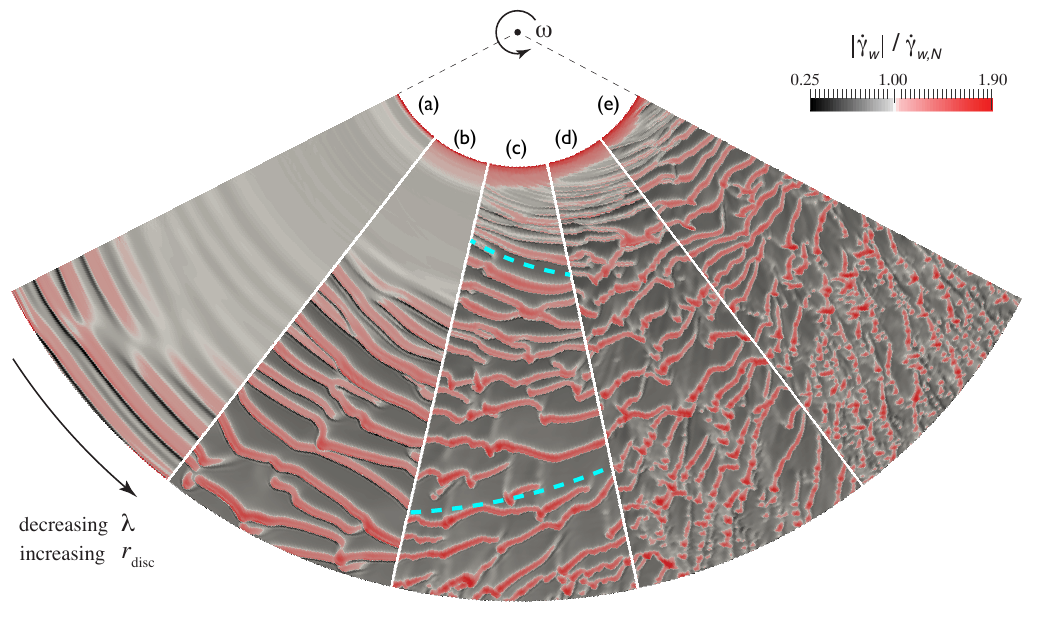}
        \caption{\label{fig:8}Wall strain rate distributions showing the local enhancement ($>1$) and reduction ($<1$) over the waveless solution (Eq. \ref{eq:gammaNlocal}). The parameters are (a) $\lambda = 0.031$, $r_{disc} = 7.81$, (b) $\lambda = 0.0156$, $r_{disc} = 9.82$, (c) $\lambda = 0.0093$, $r_{disc} = 11.68$, (d) $\lambda = 0.0068$, $r_{disc} = 12.92$, and (e) $\lambda = 0.0055$, $r_{disc} = 13.89$. Dashed lines in (c) illustrate the switching of spiral wave angles from negative to positive. }
    \end{figure*}

\begin{figure*}
        \includegraphics[width = 14cm]{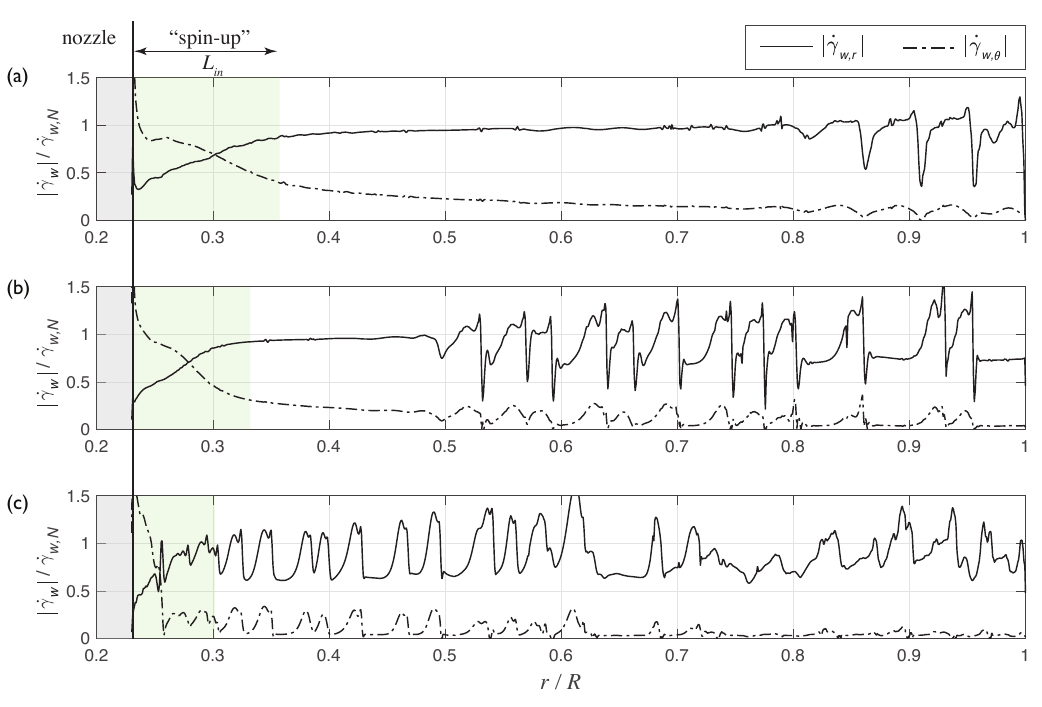}
        \caption{\label{fig:9}The radial and angular components of the wall strain rate ($\dot{\gamma}_w,r$, $\dot{\gamma}_w,\theta$) normalised by the waveless solution for (a) $\lambda = 0.031$, $r_{disc} = 7.81$, (b) $\lambda = 0.0156$, $r_{disc} = 9.82$, and (c) $\lambda = 0.0055$, $r_{disc} = 13.89$. All plots are taken at $\theta = 0.005$. The wall strain rate distributions for these parameter combinations are shown in Fig. \ref{fig:8}(a), (b), and (e), respectively. The dimensionless radial extent of the `spin-up' zone is represented by $L_{\mathrm{in}}=\left({\dot{m}^2}/{4 \pi^2 \rho_l^2 \nu_l \omega}\right)^{\frac{1}{4}}$ \cite{Butuzov1976}.}
    \end{figure*}

Confirmation that the average film height agrees with the Nusselt theory enabled us to investigate the wall shear rate distributions for the wave regimes spanning the phase diagram in Fig. \ref{fig:7} and compare against waveless flow (Eq. \ref{eq:gammaNlocal}). Figure \ref{fig:8} presents the effect of increasing inertia on the local wall shear rate. Here, we define the wall strain rate as $\dot{\gamma}_w = [\nabla \mathbf{u}+\nabla \mathbf{u}^{T}] \cdot \mathbf{n}$, where $\mathbf{n}$ is the unit vector normal to the disc surface. Smooth interface regions preceding the development of waves follow what is expected for waveless flow with $|\dot{\gamma}_w|/\dot{\gamma}_{w,N}\approx 1$ (Fig. \ref{fig:8}a,b). In the `spin-up' region close to the nozzle exit, whose dimensionless radial extent is given by 
$L_{\mathrm{in}}=\left({\dot{m}^2}/{4 \pi^2 \rho_l^2 \nu_l \omega}\right)^{\frac{1}{4}}$, 
the azimuthal component of the shear rate is at its highest as the radial flow from the nozzle experiences the surface rotation (see Fig. \ref{fig:9}). The onset of 2D waves creates alternating bands of high ($|\dot{\gamma}_w|/\dot{\gamma}_{w,N}> 1$) and low ($|\dot{\gamma}_w|/\dot{\gamma}_{w,N}< 1$) shear rates. The transition between local enhancement and reduction in shear rate is at the wavefront, where the interface curvature from the first capillary ripple leads to flow separation and a diminishing velocity gradient at the disc surface. These enhancements to the shear rate magnitude are from contributions in both the radial and azimuthal directions, with peaks in $|\dot{\gamma}_{w,r}|$ and $|\dot{\gamma}_{w,\theta}|$ exceeding $\dot{\gamma}_{w,N}$ and $\approx0.25\dot{\gamma}_{w,N}$, respectively, as shown in Fig. \ref{fig:9}. 

As $\lambda$ reduces and $r_{disc}$ increases further (see Fig. \ref{fig:8}c-e), the bands highlighted above become disordered and often breakaway coinciding with the destabilisation of the spiral wavefront and formation of 3D waves. In the case where wavelet flows emerge and dominate a significant area of the disc (see Fig. \ref{fig:8}e), we observe a diminishing $|\dot{\gamma}_{w,\theta}|$ and irregular peaks in $|\dot{\gamma}_{w,r}|$ for $r/R > 0.65$ compared to the spiral wave flows, as shown in Fig. \ref{fig:9}. This irregularity in $|\dot{\gamma}_{w,r}|$ is due in part to the 3D wave non-stationary behaviour and intermittency, which is in contrast to the stationary spiral wave flows that exist upstream. 

\begin{figure*}
        \includegraphics[width=14cm]{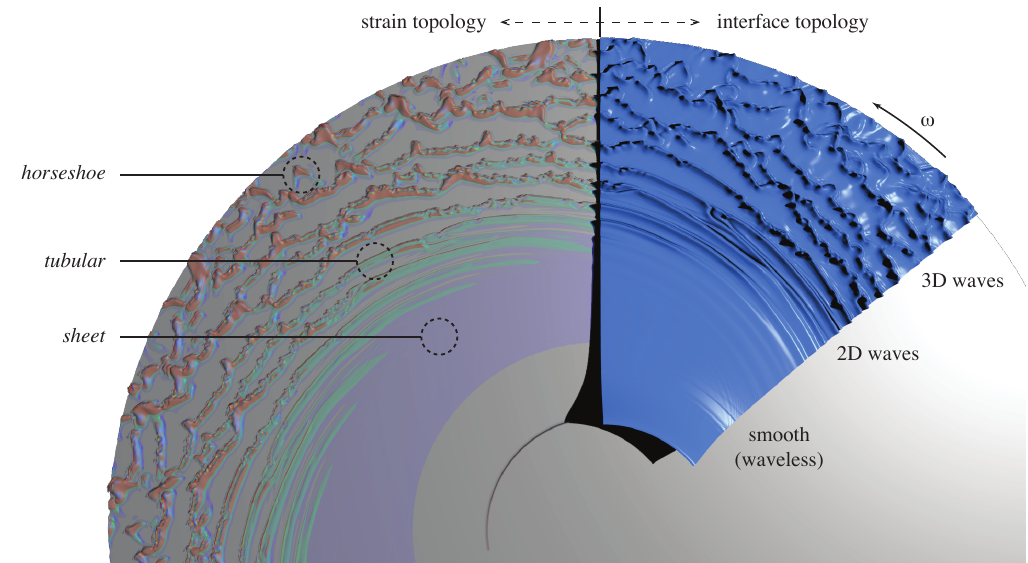}
        \caption{\label{fig:10}The influence of interface topology on that of the strain rate within the thin film. The flow parameters are $\lambda = 0.0077$, $r_{disc} = 7.07$. Strain topology is represented by isocontours of the second invariant of the rate of deformation tensor, 
        $Q_s=2 (\mathbf{S}:\mathbf{S})$ in which $\mathbf{S}=(\nabla \mathbf{u}+\nabla \mathbf{u}^{T})/2$ is the rate of deformation tensor, and $|Q_s| = 1.6, 2.2, 3.2 \times 10^6$ for blue, green, and red contours, respectively.}
    \end{figure*}

Similarly, within the film, the strain rate topology is imprinted by the wave regime as highlighted in Fig. \ref{fig:10}. Smooth or waveless flows form sheet-like structures, transition into spiral tubular ones, and then grow bulbous features along the wavefront that ultimately form isolated horseshoe structures from the breakaway of individual and groups of $\Lambda$ solitons. Changes in the magnitude and extent of these isocontours of the second invariant of the rate of deformation tensor ($Q_s$) indicate that the absolute strain rates within the main wave humps increase during the progression from 2D to 3D waves along the disc. However, the relative enhancement in wall shear rate that 3D waves have over $\gamma_2$-type spiral waves when compared to the baseline Nusselt solution is marginal. This is illustrated in the distributions of the wall shear rate magnitude presented in Fig. \ref{fig:8}, with $|\dot{\gamma}_w|<1.9\dot{\gamma}_{w,N}$ for both wave types. The radial and azimuthal components of wall shear rate presented in Fig. \ref{fig:9}c also demonstrate similar enhancements over the Nusselt solution before ($r/R < 0.6$) and after 3D waves form ($r/R > 0.6$). A key difference, however, is that the 3D waves move in the laboratory reference frame giving rise to wave-induced interfacial turbulence effects. 

\subsection{\label{sec:Evolution}Wave evolution and internal flow mechanisms}

\begin{figure*}
        \includegraphics[width=15cm]{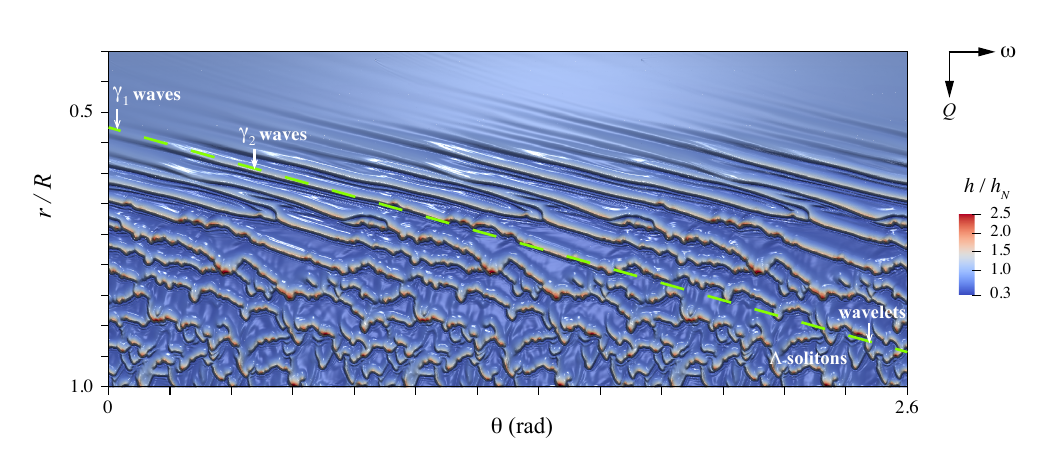}
        \caption{\label{fig:11}Interface height normalised by the Nusselt solution $h_N$ in the $r,\theta,z$ coordinate space showing the evolution of the wave structures from smooth (waveless) to 3D wavelets ($\lambda = 0.0077$, $r_{disc} = 7.07$).}
    \end{figure*}

\noindent After identifying the wave regimes, film height distributions, wall shear rates, and internal strain topology over the complete disc, in this section, we investigate the wave types and their internal dynamics in greater detail. As the phase diagram in Fig. \ref{fig:7} demonstrates, there are certain combinations of $\lambda$ and $r_{disc}$ parameters which can produce all four wave regimes on a spinning disc (\textit{waveless}, \textit{2D waves}, \textit{transition to 3D waves}, and \textit{wavelets}). These provide an opportunity to examine the evolution of the complete range of wave types observed across our experiments and simulations. We investigated the wave characteristics by unwrapping the spiral waves for the flow conditions of $\lambda = 0.0077$ and $r_{disc} = 7.07$. Transforming the interface data into the $r-\theta$ space reveals the global evolution of these waves from their inception at $r/R \approx 0.5$ ($\gamma_1$-type), growth into $\gamma_2$-type waves at $r/R \approx 0.6$, and subsequent destabilisation of the wavefront at $r/R > 0.7$ as shown in Fig. \ref{fig:11}. The spiral wave inception point also corresponds to $Ek \approx 2.5$, where Coriolis forces become important ($Ek \sim 1$). 

\begin{figure*}
        \includegraphics[width=\textwidth]{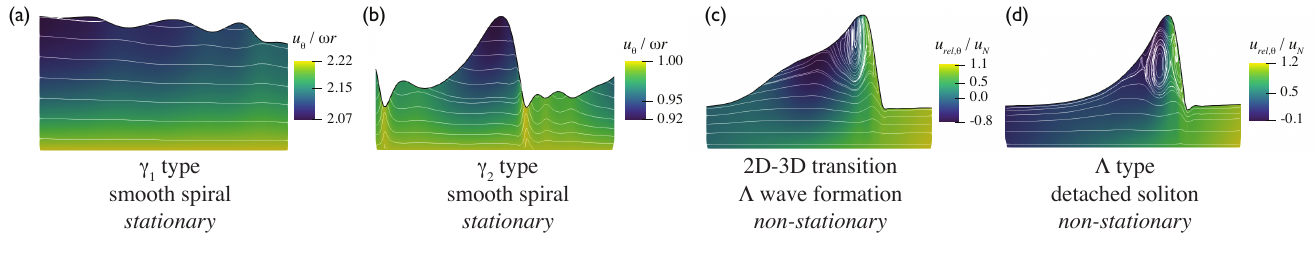}
        \caption{\label{fig:12}Flow fields in the $r-z$ plane for different wave types (a-d) that emerge from flow over a spinning disc for the same parameters as in Fig. \ref{fig:11} ($\lambda = 0.0077$, $r_{disc} = 7.07$).}
    \end{figure*}

In the region $r/D < 0.65$, the spiral wave has a smooth wavefront and is stationary in the laboratory reference frame. This creates velocity fields that are absent of internal flow circulations in the $r-z$ plane (see Fig. \ref{fig:12}a,b), unlike the falling film arrangement where the occurrence of this flow feature is dependent on the maximum velocity of the wave being equal to or greater than the wave celerity \cite{Rohlfs2015}. Here, the wave celerity is zero (stationary) in this region of the spiral wave for all cases examined in this work. The capillary ripples that form downstream with the growth of $\gamma_2$ waves over the disc leads to flow separation beneath the first minimum after the main wave hump (see Fig. \ref{fig:12}b). This separation is the result of the capillary pressure acting in the opposite direction to the radial flow, and gives rise to the sharp peaks and troughs in $|\dot\gamma_{w,r}|$ and $|\dot\gamma_{w,\theta}|$ shown previously (Fig. \ref{fig:9}). As a result, the fluid velocity beneath this first minimum is shown to be the disc speed, $u_\theta = \omega r$.   

For $r/D > 0.7$, the destabilisation of the spiral wavefront initiates the transition to 3D waves. In this region, bulbous or sawtooth structures emerge from the wavefront (see Fig. \ref{fig:5}f, Fig. \ref{fig:11}). These structures are non-stationary, moving faster than the parent spiral wave until finally detaching from its parent to form wavelets or $\Lambda$ solitons. Analysing the internal flow fields in the wave reference frame, these non-stationary waves generate flow circulations within the main wave hump (see Fig. \ref{fig:12}c,d), which appear smaller in the early stages of the wavelet formation, occupying the upper regions of the main wave hump (see Fig. \ref{fig:12}c). In the case of a completely detached $\Lambda$ soliton (see Fig. \ref{fig:12}d), this circulation occupies a greater fraction of the overall film height beneath the wave peak, $\approx0.6h_{peak}$. Interestingly, we find that these internal circulations are maintained at lower film Reynolds numbers (at least $Re = 15$) than for falling films, where the critical Reynolds number for the onset of circulations has been found to be $20 < Re < 25$ \cite{Rohlfs2015,Denner2018}.  

\begin{figure*}
        \includegraphics[width=17cm]{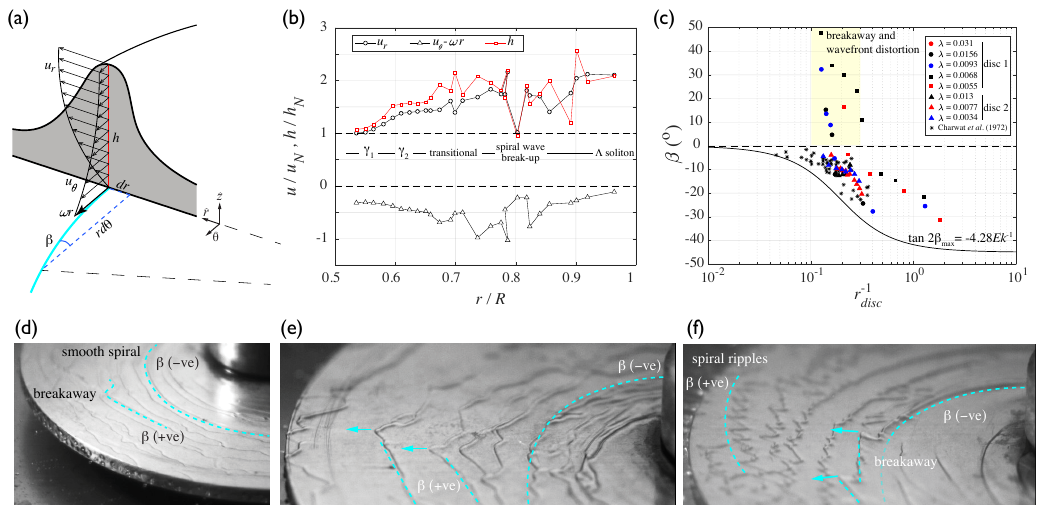}
        \caption{\label{fig:13} Spiral wave formation and transition into 3D waves. (a) Schematic of radial and azimuthal velocities ($u_r,u_{\theta}$), wave peak height ($h$), and wave angle ($\beta$). (b) Radial and azimuthal flows along a spiral wave from 2D wave inception to $\Lambda$ soliton formation ($\lambda = 0.0077$, $r_{disc} = 7.07$). (c) Spiral wave angles $\beta$ as a function of $r_{disc}$ determined from DNS across a range of $\lambda$. (d) and (e) High-speed images showing waves distorting from negative to positive $\beta$ spiral angles for $\lambda = 0.0156$, $r_{disc} = 9.82$, and (f) breakaway of 3D wave structures from a spiral wave for $\lambda = 0.0021$, $r_{disc} = 5.58$.}
    \end{figure*}

Next, we explored the change in the film velocities ($u_r,u_{\theta}$) along the spiral wave as it undergoes the transition from 2D to 3D waves (see Fig. \ref{fig:13}a,b). Sampling the depth-averaged velocity at the wave peak height (Fig. \ref{fig:13}a), we find that radial and azimuthal components increase steadily across the stationary $\gamma_1$ and $\gamma_2$ regimes. As shown previously in Fig. \ref{fig:6}, the film height is $h_{peak}\approx 1.5h_N$ in this region, reaching a plateau around $\approx 2h_N$ beyond the transitional regime. As expected, the trends in the development of film height and radial velocity are closely matched, owing to the scaling $h \sim Q^{1/3}$. In contrast, the azimuthal velocity decreases for $r/R > 0.8$, and approaches zero after a complete transition into wavelets. This drop off in azimuthal flow coincides with the spiral wave break-up event at $r/R  = 0.8$. The radial flow also drops off at this point, returning to a film velocity representative of waveless flow ($u_r \approx u_N$), and then recovering as the breakaway waves continue to flow towards the disc periphery for $r/R > 0.8$.  

The angle of the spiral waves, $\beta$ in Fig. \ref{fig:13}a, was measured for a range of hydrodynamic conditions and shown in Fig. \ref{fig:13}c and compared with other experimental data and asymptotic theory for the angle of the most unstable wave ($\tan{2\beta_{max}} = -4.28Ek^{-1}$ \cite{Charwat1972}). Most results are consistent with previous observations, with two exceptions shown. Firstly, when $\lambda < 0.01$ and $r_{disc} > 10$ the flow over the spinning disc forms waves in close proximity to the nozzle exit and within the spin-up zone (see Fig. \ref{fig:9}c). This skews the angles towards smaller (negative) values in this region of the disc (locally $0.5 < r_{disc}^{-1} < 1$) and waves accumulate with close spacings and interact with each other (see Fig. \ref{fig:8}c-e). Secondly, during the destabilisation of the spiral wavefront, we observed conditions where waves break up and spiral wave sections reconfigure their angle from $-\beta$ (unwinding in the direction of rotation) to $+\beta$. A number of examples are shown in Fig. \ref{fig:13}d-f from the experiments. We also observe the same wave-switching phenomena in our DNS results (green dashed lines, Fig. \ref{fig:8}c). Of course, this breakaway phenomena is outside the assumptions for the asymptotic theory presented by Charwat et al. \cite{Charwat1972} for spiral waves, however, overlaying the data in Fig. \ref{fig:13}c is a means of quantifying the relative changes in wave angle due to these wave-switching events. Interestingly, significant changes to the wave angle can occur during and after spiral wave breakup ($|\Delta\beta| \approx 60^{\circ}$). 

Although the unwinding of spiral waves in the direction of rotation is expected from the Coriolis effect, the mechanisms behind this switching of wave angle have yet to be elucidated and have also been observed in other high-speed imaging experiments recently \cite{Wang2023}. Analysis of our experiments and DNS results in this work reveal that the process is initiated by the formation of bulbous or sawtooth structures along the spiral wavefront. These structures grow and break away from the stationary wave, distorting the spiral wavefront and shifting the angle from negative to positive $\beta$ (Fig. \ref{fig:13}d-f). These can subsequently detach from their parent spiral wave to create $+\beta$ spiral wave sections (e.g. Fig. \ref{fig:8}c) or groups of $\Lambda$ solitons (e.g. Fig. \ref{fig:13}f) that continue downstream in the radial direction. 

\begin{figure*}
        \includegraphics[width=\textwidth]{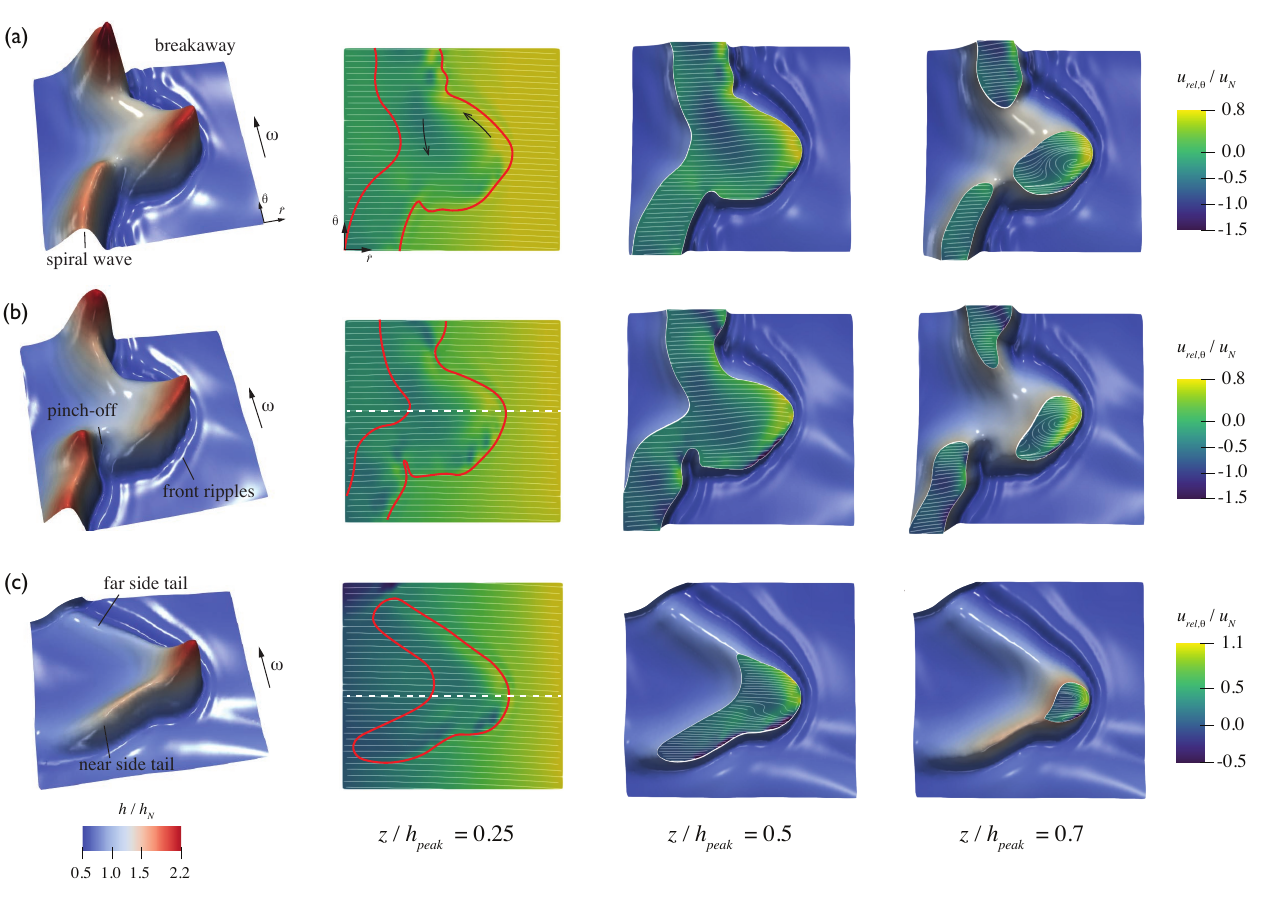}
        \caption{\label{fig:14}Flow fields in the $r-\theta$ planes during two separate wave breakaways (a,b) and for a $\Lambda$ soliton (c). All examples correspond to the conditions $\lambda = 0.0077$, $r_{disc} = 7.07$. The local $r_{disc}$, based on the main wave peak locations, are (a) $r_{disc} = 5.90$, (b) $r_{disc} = 6.13$, and (c) $r_{disc} = 6.75$. The dashed white lines in (b) and (c) coincide with the location of the $r-z$ planes presented in Fig. \ref{fig:12}c,d. The red line corresponds to the wave footprint.}
    \end{figure*}

We have also observed that the transition to 3D waves can take different pathways, from the growth and detachment of individual $\Lambda$ solitons to the breakup of spiral wave sections that reconfigure their angles. Across all scenarios, the common action is the destabilisation of the stationary spiral wavefront followed by a non-stationary wave breakaway event. We have investigated this process in further detail through examination of the interface and flow field development within the thin film. 
Figures \ref{fig:14}a,b show two separate wave breakaway events from a spiral wavefront. An asymmetric wave separation process occurs, with the first `pinch-off' from the spiral wave happening at the near side to the direction of rotation shown ($\boldsymbol{\omega}$). This asymmetry emerges as the breakaway structure extends from spiral waves that possess a non-zero angle ($\beta \neq 0$), and its growth is driven by the centrifugal force acting in the radial direction. As the wavefront of the breakaway structure adjusts to align with the centrifugal force, the near side is stretched away from the parent spiral wave. Indeed, this is a general separation process, seen for all breakaway events that occur over the spinning disc (Fig. \ref{fig:11}).

The asymmetric stretching and adjustment of the wavefront to align with the centrifugal force also influence the flow behaviour inside the film during the detachment process. Figure \ref{fig:14} presents $r-\theta$ flow fields in the wave reference frame. We find that during wave breakaway, the azimuthal flow near the far side interface moves towards the connected spiral wave, and reverses direction at the rear of the breakaway structure ($z/h_{peak} = 0.25,0.5$). This is also shown in the corresponding $r-z$ plane for the 2D-3D transition (Fig. \ref{fig:12}c). Within the main hump of the breakaway structure, the interface encloses the fluid from the near and far side spiral wave arms, resulting in the formation of a wave circulation shown in Fig. \ref{fig:14}a,b at $z/h_{peak} = 0.7$. These flow circulations in the $r-\theta$ plane rotate in the direction of the disc rotation, $\boldsymbol{\omega}$. 

Once breakaway completes and wavelets or $\Lambda$ solitons emerge (Fig. \ref{fig:14}c), the main wave hump narrows as the near- and far-side tails form, and the wavefront steepens in the $r-z$ plane (Fig. \ref{fig:12}d), and the internal circulation reduces in size in the $r-\theta$ plane. Interestingly, this detachment process generates $\Lambda$ type solitons which retain an asymmetry after breakaway. The interface height of the near-side tail was found to be consistently higher than that of the far-side tail. This arises because the near-side region contains a larger volume of liquid during the breakaway process and is illustrated in Fig. \ref{fig:14}b ($z/h_{peak} = 0.25$) by the difference in the wave footprint (red outline) above and below the constant $\theta$ line that has been plotted through the wavefront. The subsequent formation of the $\Lambda$ soliton retains the additional fluid volume within the tail, increasing the interface height above that of the far-side tail. For large $r_{disc}$ (or $Ek$), it is probable that this asymmetry will diminish as the angle of the spiral wave approaches $\beta \approx 0$ for $r_{disc} \gg 100$ (Fig. \ref{fig:13}c). Such large $r_{disc}$ conditions become comparable with the hydrodynamics of falling films, as shown by the applicability of integral boundary layer solutions to the falling film problem in the large $Ek$ limit \cite{Sisoev2003}.

\section{Conclusions \label{sec:conclusions}}

\noindent The flow of a thin liquid film over a spinning disc has been investigated by means of experiments and three-dimensional (3D) direct numerical simulations. Four different wave regimes were observed which spanned smooth (waveless), two-dimensional (2D) spiral waves, transition to 3D waves, and finally 3D waves or wavelets. Using two governing dimensionless parameters to describe the hydrodynamics of spinning disc flows, which correspond to an inverse Weber ($\lambda$) and a modified Ekman ($r_{disc}$) number, a phase map was constructed to classify these wave regimes and identify the transition boundaries. A comprehensive examination of the nonlinear evolution of waves across these transitional boundaries, which consist of $\gamma_1$, $\gamma_2$, and $\Lambda$-type wave families, was used to reveal unique hydrodynamic characteristics of inertia-dominated thin film flows over spinning discs. 

The formation of $\gamma_2$-type spiral waves and $\Lambda$ type 3D waves creates local peaks in wall strain rate that are $\approx 1.5-2$ times greater than that associated with the waveless solution. Stationary spiral waves emerge from the Coriolis effect followed by a destabilisation of their wavefronts and a transition to non-stationary wave structures. This leads to wave sections breaking away from the parent spiral wave and also the creation of individual wavelets. These non-stationary waves are asymmetric, and through stretching and realignment with the centrifugal force during formation, create internal flow circulations in both the radial-axial and radial-azimuthal directions. These circulations were also found to exist at lower film Reynolds numbers than previously identified for falling films. 

Practically, our findings can be used to determine operating parameters for wave selectivity and explain the underlying mechanisms that lead to the efficient mixing environment observed in chemical reactors which enhances heat and mass transport phenomena. Future work will focus on higher film Reynolds numbers for which the flow regime within the film transitions to turbulence. 

\begin{acknowledgments}
\noindent J. S. acknowledges funding from the European Union's Horizon 2020 research and innovation programme under the Marie Sk\l{}odowska-Curie Individual Fellowship grant agreement No. 707340. O. K. M. acknowledges funding from PETRONAS and the Royal Academy of Engineering for a Research Chair in Multiphase Fluid Dynamics, and from the Engineering and Physical Sciences Research Council UK through the MEMPHIS (EP/K003976/1) and PREMIERE (EP/T000414/1) Programme Grants. 
\end{acknowledgments}

\bibliography{main}

\end{document}